\documentclass[pre,aps,onecolumn,showpcs,floatfix,amsmath,amssymb]{revtex4}
\usepackage{amsmath}
\usepackage{graphicx,dcolumn,bm,natbib}
\usepackage{bm}        
\usepackage{mathrsfs} 
\usepackage{color}

\def\3dots{\:\raisebox{-0.5ex}{$\stackrel{\textstyle.}{:}$}\:}

\def\beq{\begin{equation}}
\def\eeq{\end{equation}}
\def\bea{\begin{eqnarray}}
\def\eea{\end{eqnarray}}

\begin{document}
\title{Active elastic dimers: Cells moving on rigid tracks}
\author{J. H. Lopez$^1$, Moumita Das$^2$, and J. M. Schwarz$^1$}
\affiliation{$^1$Department of Physics, Syracuse University, Syracuse, NY 13244, USA}
\affiliation{$^2$School of Physics and Astronomy, Rochester Institute of Technology, Rochester, NY 14623, USA}

\begin{abstract}
Experiments suggest that the migration of some cells in the three-dimensional extra cellular matrix bears strong resemblance to one-dimensional cell migration. Motivated by this observation, we construct and study a minimal one-dimensional model cell made of two beads and an active spring moving along a rigid track.   The active spring models the stress fibers with their myosin-driven contractility and alpha-actinin-driven extendability, while the friction coefficients of the two beads describe the catch/slip bond behavior of the integrins in focal adhesions. In the absence of active noise, net motion arises from an interplay between active contractility (and passive extendability) of the stress fibers and an asymmetry between the front and back of the cell due to catch bond behavior of integrins at the front of the cell and slip bond behavior of integrins at the back. We obtain reasonable cell speeds with independently estimated parameters. We also study the effects of hysteresis in the active spring, due to catch bond behavior and the dynamics of cross-linking, and the addition of active noise on the motion of the cell.  Our model highlights the role of alpha-actinin in three-dimensional cell motility and does not require Arp2/3 actin filament nucleation for net motion. 
\end{abstract}

\maketitle

\section{Introduction}

Epithelial cells crawl to heal a wound, white blood cells migrate to chase and ingest harmful bacteria, and, in an embryo, neural crest cells move away from the neural tube to generate neurons, bone cells, and muscle cells~\cite{alberts,bray}. Since cell motility is integral to a wide range of physiological processes, quantitative understanding of it is an important step in the quantification of cell biology at and beyond the cell size scale.  
    
To date, most quantitative understanding of cell motility pertains to cells crawling on surfaces~\cite{mogilner}. For example, one can predict the shape of a crawling cell based on its speed~\cite{keren}. And yet, is a smooth surface a native environment for a crawling cell? The answer is typically no. For instance, epithelial cells must crawl through the three-dimensional extracellular matrix (ECM) to heal a wound.  The ECM consists mostly of fibrous collagen with a pore-size that can range up to the order of the cell size (tens of microns)~\cite{wolf}. So how does this type of environment affect single cell motility in terms of speed, overall direction of migration, and sensitivity or robustness to changes in the environment? 

There has been a recent explosion in experiments tackling this question [5-17]. These experiments clearly demonstrate that cells crawling through the ECM can take on a very different shape from the ones crawling in two dimensions, namely, they mimic the fibrous environment of the ECM by elongating as they traverse along fibers~\cite{yamada}.  An elongated shape is very different from the fan-like cell shapes observed in two dimensions such that new approaches to quantitative modeling may be needed. Based on these results, cell crawling experiments in one dimension have been conducted to study how one-dimensional single cell migration compares to three-dimensional single cell migration along fibers~\cite{kumar,king}. Moreover, as the cell crawls through the ECM, the cell remodels it, again, calling for new approaches to prior two-dimensional quantitative modeling. While three-dimensional cell migration experiments are becoming numerous, there have been very few studies focused on quantitative modeling of these experiments.
 
Here, as a first step, we focus on modeling cells that move along very taut ECM fibers---taut enough such that they are essentially featureless (rigid) tracks. To do so, we build a one-dimensional model of cell motility along one fiber, or track, via a bead-active-spring model, the properties of which will be described below. See Figure 1. 
Bead-spring models have been successfully used to elucidate the role of cell mechanical properties in driving shape dynamics for cells crawling in two dimensions. In particular, Refs. [21] and [22] have captured bipedal locomotion in crawling cells using a two dimensional bead-spring model. Ref. [23] introduces a one-dimensional Brownian inchworm model for directed self-propulsion in the presence of noise. This model consists of an elastic dimer representing the front and rear of the self-propelled particle and shows that an effective friction force that depends on the elastic coupling between the two beads can rectify diffusive motion to lead to directed motion (even in the absence of an externally imposed gradient). 

In our bead-active spring model, the spring represent stress fibers comprised of actin, myosin, and cross-linker complexes~\cite{fiber}.   Because the stress fibers contain myosin motors, they contain an ``active'' component.  ATP-driven myosin walk towards the plus end of the actin filament such that two actin filaments of opposite orientation coupled via myosin will contract, as in muscle. While the orientation of the actin filaments is not as regular as in muscle, i.e. some filaments coupled via myosin are not oppositely oriented, overall contraction is still occurs~\cite{stressfibercontracts}. So the spring denotes the stress fibers, and the beads denote the location of focal adhesions, which enable the stress fibers connect to the ECM.  Integrins are one of the main proteins comprising focal adhesions~\cite{waterman}. As far as the type of molecular bonding, it has been shown that integrins can act as catch bonds~\cite{kong}.  For catch bonds, the bond lifetime increases with increasing force before decreasing with even further increase in force, while for slip bonds, the bond lifetime decreases with increasing force~\cite{catch,bellmodel}. Catch bond behavior is less intuitive than slip bond, but their enhanced strength over a range of forces may play a key role in how cells respond to and explore their mechanical environment. 

With these minimal ingredients in our quantitative model, we explore the following questions: What is the interplay between the kinetics of focal adhesion binding to the rigid track and the active mechanics of the stress fibers in affecting cell speed in this constrained environment?  What about the role of myosin (active cross-linkers) versus passive cross-linkers in one-dimensional cell crawling? Also, what is the role of randomness, due to activity, on cell crawling?  More precisely, how robust is the motion to randomness? The answers to these questions can then be tested {\it in vitro} with various knockdowns and/or mutant fibroblasts, for example, crawling along fabricated microbridges (with no side walls) as a starting point for understanding how a cell moves in the complicated microenvironment of the ECM.

The organization of the paper is as follows. The next section details the ingredients for the minimal bead-active spring model along with the equations of motion of the model. Section III presents estimates of the parameters used. Section IV explores solutions to these equations, i.e. cell movement, in the physiological part of the parameter space.  The final section, Section V, addresses the implications of our work.

\begin{figure}
\begin{center}
\includegraphics[width=7cm]{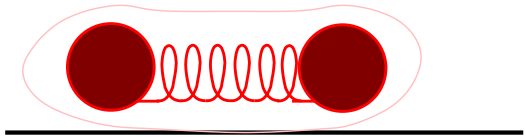}
\caption{Side-view schematic of a two-bead-spring model of a cell crawling along a narrow track.}
\end{center}
\end{figure}

\section{A minimal model}

We start by asking the following question: which aspects of two-dimensional cell movement hold for cells crawling along the fibers of the ECM, one of the native environments for a crawling cell? 
Two-dimensional cell crawling studies support the following scenario~\cite{mogilner}: The cell extends its front via actin filament nucleation and polymerization and then creates mature focal adhesions under the new extension. Meanwhile, focal adhesions are disassembled near the rear of the cell so that the rear can retract to catch up with the front, which has since continued to extend. The retraction is myosin-driven since the use of blebbistatin suppresses motility of a cell~\cite{cellcrawl}, though leading edge cell fragments can continue to move via actin-treadmilling~\cite{cellfrag}. In this two-dimensional scenario, actin filament nucleation is driven by the branching agent, Arp2/3~\cite{arp1,arp2}. Arp2/3 nucleates branched filaments at a reasonably regular angle of 70 degree from the polymerizing end of actin filaments and, therefore, helps set the lateral extent of the leading edge of the crawling cell. This extent can be broad for cells crawling on two-dimensional substrates, resulting in fan-like shapes at the leading edge.

 Some aspects of this description of two-dimensional cell crawling still hold for cell migration on ECM fibers, in the sense that there is extension, the assembly and disassembly of focal adhesions, and contractility driven by myosin. The most notable difference from two-dimensional studies is the elongated shape of cells undergoing mesenchymal migration, or crawling along fibers. This observation has led researchers to conjecture that this particular mode of cell migration is effectively one-dimensional migration~\cite{yamada}. There are other observations that are consistent with the conjecture.  For instance, Arp2/3 does not appear to be as important in generating motion here since the rather wide branch angle leads to large lateral lengths, which would not be commensurate with the underlying fiber~\cite{anjil}.  Instead, actin filament nucleation via Arp2/3 is important for generating pseudopods whose possible function could be to search out for other ECM fibers to move along. 

Here we study the motion along one fiber only, and focus on the interplay between stress fibers and focal adhesion. To quantify the interplay between focal adhesions and myosin-driven contractility, we construct a minimal one-dimensional model for a crawling cell as two beads connected by an active spring. The two beads denote the two ends of a cell that attach to the surface via focal adhesions. While focal adhesions occur throughout the cell, traction force microscopy indicates that the focal adhesions exert the largest stresses at the edges of a crawling cell on surfaces~\cite{dembo}. We assume that the same observation holds for cells crawling in confined constrictions. Bead 1, denoted by position $x_{1}(t)$, is to the right of Bead 2, denoted by $x_{2}(t)$ as shown in Fig. 1. The beads have masses $m_1$ and $m_2$, and friction coefficients $\gamma_1$ and $\gamma_2$ respectively. The friction coefficients model the focal adhesions, or attachment to the fiber, while the active spring in-between the two beads denotes the stress fibers. Let us now quantify the concept of an active spring.

\subsection{Stress fibers as active springs with two equilibrium lengths}

Stress fibers primarily consist of actin filaments, myosin, and alpha-actinin, a passive cross-linker~\cite{fiber}. A few other proteins, such as zxyin, colocalizes with alpha-actinin~\cite{zxyin}. The stress fiber is made up of parallel arrangements of actomyosin units in series.  Each actomyosin unit is considered as two actin filament rods connected by a myosin minifilament and alpha-actinin at each end.  Since the stress fibers in cells crawling in constrained geometries exhibit more ordered stress fibers than the cells crawling on surfaces, using this fundamental muscle-like element is very useful~\cite{friedl2}.  See Figure 2. For a static cell, the stress fiber is under contractile tension as it adheres to the substrate.  In a moving cell, the focal adhesions are being created and destroyed. Since myosin exhibit catch bond behaviour with an optimum load force of about 6 pN per motor, the myosin may not always be under sufficient load (or too much load) to walk efficiently along the actin filaments~\cite{guilford}.

More specifically, when focal adhesions are just beginning to form at the front of the cell, myosin are not pulling due to the small applied load. And when myosin are not pulling, the plus end of actin filaments separate/extend.  We argue that the plus ends extend to relieve the strain in the alpha-actinin such that it approaches its equilibrium configuration. See Figure 2.  In this alpha-actinin extension mode, the mechanical stiffness of the active spring, $k$, is primarily due to the stiffness of the alpha-actinin. Moreover, the equilibrium spring length of the active spring is denoted by $x_{eq1}$. As the focal adhesions at the front of the cell mature over a time scale of seconds~\cite{fagrowth}, the myosin come under load again such that they ``catch'' and exert contractile forces on each pair of actin filaments to induce a contracted mode causing the alpha-actinin to stretch and rotate in the opposite direction. In this mode, myosin provide the mechanical stiffness of the spring and there is a second equilibrium spring length, $x_{eq1}-x_{eq2}$, with $x_{eq2}<x_{eq1}$ as indicated by the isolated stress fiber experiments~\cite{stressfibercontracts}. 

How then does the stress fiber switch back the extension mode? As the myosin contract, strain builds in the alpha-actinin. This strain build-up can be enhanced by zxyin binding to the alpha-actinin such that the myosin no longer ``catch'' and a transition is then made to the extending mode. Experiments tracking zyxin in static cells find that it colocalizes to places along the stress fiber under high tension and have argued that zyxin could act as some molecular switch from one mechanical state to another~\cite{colombelli}. 

Given these two modes of the stress fiber, passive extension and active (motor) contraction, {\em we model the elasticity of the stress fiber as a spring with two different equilibrium spring lengths}. The transition between the two modes of the active spring is determined by the extension of the spring. The larger the extension of the spring, the more tensile load on the myosin so as to induce contractility of the myosin. Therefore, a simple model for the equilibrium spring length, $x_{eq}$, of this active spring is
\begin{equation}
x_{eq}=x_{eq1}-x_{eq2}\Theta(x_1-x_2-l),
\end{equation}
where $\Theta(x_1-x_2-l)$ is the Heaviside step function. See Figure 3. With this choice, when $x_1-x_2>l$, the equilibrium spring length is shorter when myosin actively pull and longer when the myosin do not.  Moreover, $l$ is bounded below by $x_{eq1}-x_{eq2}$ and above by $x_{eq1}$. With this changing equilibrium spring length, the spring is now an active contractile element.  

In addition to the catch-bond kinetics of the acto-myosin bonds, alpha-actinin exhibits catch-bond kinetics as well~\cite{yao}. Catch-bond kinetics indicate some sort of conformational change in the protein such that the conformation of the alpha-actinin in the extended mode may indeed be different than when in the contracting mode.  The binding of zxyin may also affect the conformation of the alpha-actinin. A possible change in conformation of the alpha-actinin suggests that the transition between extension and contraction is not necessarily reversible, particularly if zxyin bind in one conformation (but not the other)~\cite{colombelli}. Moreover, when the active spring is in its extended mode, there is less overlap between the actin filaments such that it is less likely that additional alpha-actinin can bind together two actin filaments. Conversely, when the active spring is in its contracted state, it is more likely that an additional alpha-actinin can link two actin filaments together.  Therefore, for the active spring to extend, it must overcome the additional binding energy of the added alpha-actinin, i.e. bonds must be broken. However, this additional binding energy is not present as the active spring contracts. 

To account for potential conformational changes in the alpha-actinin, additional alpha-actinin binding, and even internal frictional losses, we allow $l$ to take on two values, $l^{\uparrow}$, as the active spring extends and $l^{\downarrow}$ as the active spring compresses with $l^{\uparrow}>l^{\downarrow}$. In sum, the equilibrium active spring length takes on the form,
\begin{equation}
x_{eq}=x_{eq1}-x_{eq2}\Theta(x_1-x_2-l^{\uparrow}),
\end{equation}
when the active spring is extending and
\begin{equation}
x_{eq}=x_{eq1}-x_{eq2}\Theta(x_1-x_2-l^{\downarrow}),
\end{equation}
when the active spring is contracting. This means that the description for $x_{eq}$ contains {\em hysteresis}. Such hysteresis in stress-strain behavior is often found in 
materials where the strain history affects the observed stress giving rise to different stress-strain paths for loading and unloading. Prime examples are the phenomenological Johnson-Segalman model 
of viscoelastic behavior~\cite{johnson-segalman}, and the experimentally observed strain history dependent mechanical response of soft biological tissue~\cite{munster-pnas}. We must also point out that a recent viscoelastic model for stress fibers is an active version of an viscoelastic polymer model~\cite{schwarz}. Because the width of the hysteresis represents a strain barrier and the height a strain ``input'', the height of the hysteresis loop must be greater than the hysteresis width to generate motion.

\begin{figure}
\begin{center}
\includegraphics[width=9cm]{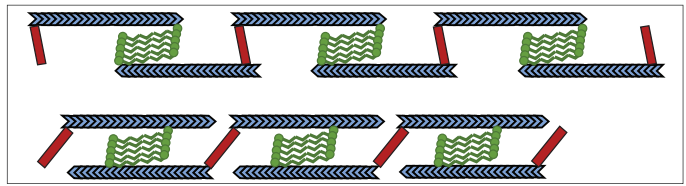}
\caption{Schematic of contractile units in a stress fiber in extended mode (top) and contracted mode (bottom).  The blue filaments represent actin filaments, red rectangles, alpha-actinin, and the green shapes, myosin minifilaments.  For simplicity, we have not shown any contractile units in parallel, only three units in series.}
\end{center}
\end{figure}

\subsection{Focal adhesions provide an elastic friction}

Now that we have quantified our active spring, we turn to the focal adhesions. The mechanical interaction between the migrating cell and the ECM are mediated by cell surface receptors and associated ligands in the ECM.  The ECM glycoprotein fibronectin and the transmembrane receptor proteins of the integrin family, form the major and most well-characterized receptor-ligand pair~\cite{integrin}.  In their inactive state, integrins exist in a bent, relaxed form so as to avoid the formation of physiologically harmful cell-cell or cell-ECM connections. Once they are activated via a vertical load, they undergo a conformational change to an extended state~\cite{kong,sheetz}.  When in this state, AFM experiments find that integrins respond additionally to an increase in the lateral distance between the two extended dimers with an increased bond lifetimes for applied forces up to 30 pN~\cite{kong}. In other words, integrin can act as a catch bond. It may indeed be the maturation of the focal adhesion that triggers this lateral distance and, thereby, the catch bond mechanism of the integrins~\cite{sheetz}. 

In light of these findings, we conjecture that in the front of the cell, integrins are more likely to act as catch bonds due to maturation of focal adhesions.  In the back of the cell, however, integrin act as typical slip bonds, where focal adhesions are merely being disassembled. Therefore, in the front of the cell, the initiation of focal adhesions call for a ``small'' friction coefficient, but once the focal adhesion forms and develops, it has a large friction coefficient when compared to an integrin slip bond. This ``catching'' mechanism of cell-track adhesion allows the cell's front to expand and explore new territory and after having done that, then allows for the cell's rear to retract with the cell front not losing grip on the new territory it just explored due to the catch bond mechanism. Since the stress fibers and the focal adhesions are connected, we define  
\begin{equation}
\gamma_1=\gamma_{11}+\gamma_{12}\Theta(x_1-x_2-l^{\uparrow(\downarrow)})
\end{equation}
with $\gamma_{11},\gamma_{12}>0$ and $\gamma_{11}<\gamma_{12}$. For small extensions of the cell, the friction at the leading bead is smaller than for large extensions. Larger friction implies a larger unbinding rate for integrins and, therefore, the integrins can more effectively grip the track. In addition, because the integrins track the myosin activity, the hysteresis exhibited by the myosin is also exhibited in the friction. See Figure 3. Finally, $\gamma_{2}$, the friction coefficient for the now ``rear'' bead, is assumed to be constant with the integrins acting as ordinary slip bonds.

\subsection{Equations of motion}

With the stress fibers modeled as an active spring with spring constant, $k$, and a changing equilibrium spring length, and the focal adhesions localized at tshe front and the back beads of the two bead-active spring model, the two coupled equations for the motion of the beads are as follows:
\begin{equation} \label{model}
m_i\ddot{x}_{i}(t) + \gamma_{i}(x_1,x_2,l^{\uparrow},l^{\downarrow})\dot{x}_{i}(t) = \pm k(x_1-x_2-x_{eq}(x_1,x_2,l^{\uparrow},l^{\downarrow})) +\sqrt{A_i}\zeta_i(t).
\end{equation}
Note that we have included an ``active noise'' term, where $A_i$ is the variance of the active noise contribution due to stochasticity in motor activity, and $\zeta_i(t)$ is a Gaussian random variable with $<\zeta_i(t)>=0$ and $<\zeta_i(t)\zeta_j(t')>=\delta_{ij}\delta(t-t')$. Here, $A_i$ does not satisfy a fluctuation-dissipation relation and is not associated with any temperature. We will study this model for both $A_i=0$ (deterministic) and $A_i>0$ (non-deterministic).

\begin{figure}
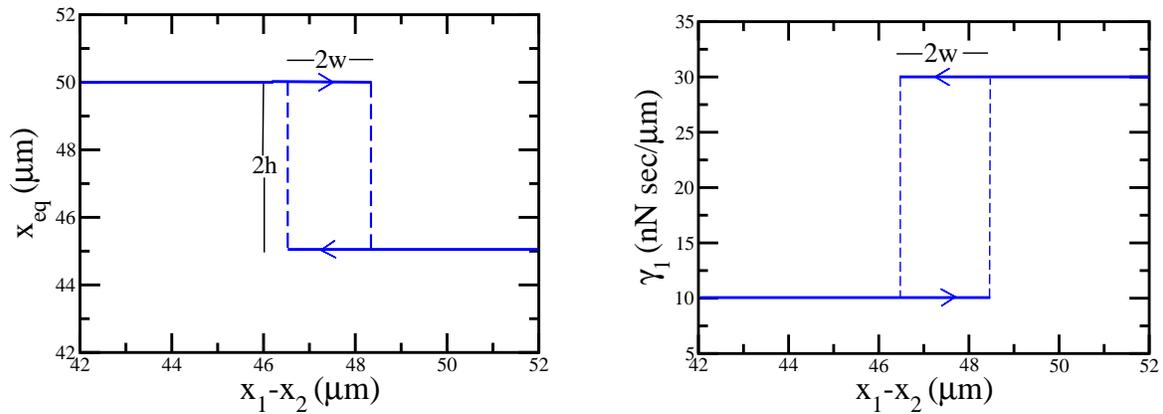

\begin{center}
\includegraphics[width=7.2cm]{hysteresis.final.eps}\hspace{1cm}
\includegraphics[width=7cm]{hysteresis2.final.eps}
\caption{Left (a): Plot of the equilibrium spring length $x_{eq}$ as a function of $x_1-x_2$. Right (b): Plot of friction coefficient $\gamma_1$ as a function of $x_1-x_2$. The parameters used are listed in Table I.}
\label{hysteresis}
\end{center}
\end{figure}

\section{Estimation of parameters}

Now that we have the formal solutions for the relative and center-of-mass coordinates, let us present estimates for the parameters involved before analyzing the solutions in further detail. 

\subsection{Active spring parameters}

The actomyosin units account for both the passive mechanical stiffness and the active contractile properties of the stress fiber.  The stiffness of the myosin minifilament is represented by a spring of stiffness $N_m k_{m}$, where $N_m$ is the number of myosin motors in the minifilament and $k_m$ is the spring constant for each individual myosin with $k_m\approx 1 \,\,pN/nm$ ($1 \,\, pN/nm = 1 \,\, nN/\mu m$) and $N_m\approx 50$~\cite{gardel}. For $N_m\approx 50$, the typical length of a myosin minifilament is $0.3 \,\, \mu m$, while its width is approximately 30 nm~\cite{pollardmyosin}, which is also consistent with the approximate length of alpha-actinin. Each motor exerts equal and opposite contractile forces on the two actin filaments, denoted each by $f$, on the two actin filaments. Each myosin motor head can exert a maximum of $f/2=3\,\, pN$ of contractile force~\cite{howard}.  The actin filaments are modeled as rigid filaments with the pair of spanning a maximum length $L$. Typically, $L=1 \,\,\mu m$. Each alpha-actinin is modeled as a linear spring with spring constant, $k_0\approx 50\,\,pN/nm$ and rest length $L_a$ that can change due to potential conformational changes in the alpha-actinin between the extending and contracting modes of the actomyosin units~\cite{soncini}.   

As mentioned previously, experiments on isolated stress fibers find up to a 23 percent decrease in length with the addition of ATP~\cite{stressfibercontracts}. In the extended mode, we use an equilibrium spring length, $x_{1eq}=50$ microns since stress fibers typically consist of about 50 actomyosin units in series and each of the units span a maximum of 1 micron~\cite{length}. Given the experimental results for percentage of decrease in length of the stress fiber due to myosin contractility, we will explore a range of percentages around 10 percent. 

With the above ingredients, we can also estimate the effective stiffness of the stress fiber active-spring as follows.  The effective stiffness of a myosin minifilament consisting of $N_m\approx 50$ myosin motors, each with a myosin spring constant approximately $1 \,\,pN/nm$ in parallel, is $50\,\, pN/nm$. In the extended mode of the active spring, the alpha-actinin contributes to the elasticity, in the contracted mode, the myosin minifilaments contribute to the stress fiber elasticity. This leads to a spring stiffness of $\sim 50\,\, pN/nm$ for either mode for each unit such that $k=50 \,pN/nm\,\,(N_p/N_s)$, where $N_s$ is the number of acto-myosin contractile units in series and $N_p$ in parallel. With $N_s=50$ and $N_p=1$, $k\approx 1\,\, pN/nm$. For $N_p>$1, the effective stress fiber spring constant is larger.

\subsection{Friction parameters} 

We model the integrins as springs with dissociation kinetics described by catch or slip bond behavior. Each integrin bond can be thought of as a single Hookean spring and allowed to fail at one point at the cell-ECM interface.  At the back of the cell, the unbinding kinetics of the integrin bond will follow slip bond behavior with an effective dissociation rate, $K_{off}^*$, that increases exponentially according to a Bell Model~\cite{bellmodel}, or 

\beq
K_{off}^*=K_{off} e^{F_{bond}/F_b}
\eeq
where $K_{off}$ is the unforced dissociation rate of the slip bond, $F_b= k_B T/ \psi$ is the characteristic bond rupture force and $\psi$ is a characteristic unbinding lengthscale, and $F_{bond}$ is the tension within an individual slip bond spring.  Hence, the slip bond lifetime simply decreases with increasing applied tensile force.

For the front bead, the integrin bond acts as a catch bond in the presence of developing focal adhesions and the dissociation kinetics is a sum of two pathways---one where the bond is strengthened by the applied force and other where it is weakened.  The summative unbinding rate can be written as follows:
\beq
K_{off}^*= K_s e^{F_{bond}/F_b} +   K_c e^{-F_{bond}/F_b}
\eeq
where, the unforced unbinding rates $K_s=K_{off} e^{-F_s/F_b}$ and $K_c= K_{off} e^{F_c/F_b}$ are each associated with each pathway~\cite{storm}.

Once $K_{off}$ is known, the friction coefficients can be computed using using the formula,
\begin{equation}
\gamma=\frac{N_{int}k_{int}}{K_{off}^*},
\end{equation}
where $N_{int}$ is the number of bound integrins and $k_{int}$ is the spring constant of the molecular bond. We use $k_{int}\approx 10\,\,pN/nm$ and $N_{int}\approx 1$, though we will explore other values. Since integrins form the bond between the cell and the substrate, we use the kinetic curve obtained from Kong and collaborators for the lifetime of a single bond as a function of applied loard~\cite{kong}.  For the front bead, we use $K_{off}^*= 1\,\,s^{-1}$ to compute $\gamma_{11}=10\,\,nN\,s/\mu m$, the weaker coefficient, and an off-rate of 1/3 inverse seconds for the stronger value of the friction coefficient of the front bead, leading to $\gamma_{12}=20\,\,nN\,s/\mu m$. Then, $\gamma_{11}+\gamma_{12}=30\,\,nN\,s/\mu m$. For the back bead, $K_c=0$ and we use $K_{off}^*=0.5\,\,s^{-1}$ to arrive at $\gamma_2=20\,\,nN\,s/\mu m$.

\begin{table}[h]
\setlength{\tabcolsep}{12pt}
\centering
\begin{tabular}{|c| c|}
\hline
Parameters & Values\\
\hline
$k$ & $1\,\, nN/\mu m$ \\ 
$x_{eq1}$ & $50 \,\,\mu m$\\ 
$x_{eq2}$ & $5\,\,\mu m$ \\
$l^{\downarrow}$ & $\sim 46.5\,\,\mu m$\\  
$l^{\uparrow}$ & $\sim 48.5\,\,\mu m$\\
$\gamma_{11}$ & $10\,\,nN\, s/\mu m$\\
$\gamma_{12}$ & $20\,\,nN\, s/\mu m$\\
$\gamma_{2}$ & $20\,\,nN \,s/\mu m$\\
$m_1,m_2$ & $\sim 0$\\
$A_1,A_2$ & $\sim 0$\\
\hline
\end{tabular}
\caption{Table of parameters used.} 
\end{table}

\section{Results}

To solve the equations of motion, (Eq. 5), we neglect inertia, as demanded by the physiological conditions. We then first investigate the cell crawler in the absence of any noise such that $A_1=A_2=0$. Next, defining $x=x_1-x_2$ and subtracting the equation of motion for $x_2$ from $x_1$, we arrive at
\begin{equation}
\dot{x}=-(\frac{1}{\gamma_1(x,l^{\uparrow(\downarrow)})}+\frac{1}{\gamma_2})k(x-x_{eq}(x,l^{\uparrow(\downarrow)})),
\end{equation}
depending on whether the spring is extending or compressing. Similarly, the equation of motion for the center of mass is 
\begin{equation}
v_{cm}(t)=\dot{x}_{cm}=-\frac{1}{2}(\frac{1}{\gamma_1(x,l^{\uparrow(\downarrow)})}-\frac{1}{\gamma_2})k(x-x_{eq}(x,l^{\uparrow(\downarrow)})),
\end{equation}
where $x_{cm}=\frac{x_1+x_2}{2}$. A non-zero center of mass velocity translates to motion of the cell. 

Since the center of mass velocity equation depends on $x$, we first solve the equation of motion for $x$.  To do so, we break up the system into when the equilibrium spring length is $x_{eq1}$ and when the equilibrium spring length is $x_{eq1}-x_{eq2}$.  In the former case, 
\begin{equation}
x_{I}(t)=x_{eq1}+(x(0)-x_{eq1})e^{-\frac{k}{\gamma_2}\frac{(\gamma_{11}+\gamma_2)}{(\gamma_{11})}t}, 
\end{equation}  
and in the latter, 
\begin{equation}
x_{II}(t)=x_{eq1}-x_{eq2}+(x(0)-x_{eq1}+x_{eq2})e^{-\frac{k}{\gamma_2}\frac{(\gamma_{11}+\gamma_{12}+\gamma_2)}{(\gamma_{11}+\gamma_{12})}t}.
\end{equation}

Now, depending on the history of the spring, be it contracting or extending, we can piece together these solutions accordingly. For example, assume $x(0)\ge l^{\uparrow}$, then $x$ decreases and obeys $x_{II}(t)$, which decreases exponentially with time.  This is because the cell has ``over-extended itself'' in its search for new territory and now the focal adhesions have matured so both the equilibrium spring length is decreased, due to myosin-induced contractility, and the front catch bonds ``catch'' such that the back of the cell can catch up with the front without losing new ground. After the initial decrease in $x$, as soon as $x$ decreases below $l^{\downarrow}$, then the myosin effectively stop pulling, due to strain built up in the stress fibers from the focal adhesions and alpha-actinin, and the equilibrium spring length increases with new focal adhesions developing at the front. Once this happens, we re-zero our time clock back to $t=0$ and iterate $x_{I}(t)$, an exponential expansion given the initial condition, until $x$ becomes larger than $l^{\uparrow}$ such that $x_{II}(t)$ solutions become valid and the process repeats itself. As we will see below, this cyclic process in an overdamped system leads to net motion due to (1) the switching between the two equilibrium spring constants, which drives the overdamped system out-of-equilibrium, and (2) the asymmetry in the friction coefficients.  Both properties are needed for motion.

Let us analyze the active dimer motion as a function of the width and height of the hysteresis loop. Defining $w= \frac{1}{2}(l^\uparrow - l^\downarrow)$ and $h=\frac{1}{2}x_{eq2}$, the two timescales over which the cell undergoes extension and contraction are given by $t_I= \beta \log{\frac{h+w}{h-w}}$ 
and $t_{II}= \alpha \log{\frac{h+w}{h-w}}$ respectively, where $\alpha= \gamma_2 (\gamma_{11} + \gamma_{12})/k(\gamma_{11} + \gamma_{12}+\gamma_{2})$ and 
$\beta= \gamma_2 \gamma_{11}/k(\gamma_{11} + \gamma_{2})$. As stated earlier, $w<h$ for motion to occur since the active strain energy generated by the myosin must overcome the strain barrier by the alpha-actinin. When the active dimer is extending to relieve the strain in the alpha-actinin and $x>l^\downarrow$, the maximum and minimum values of the center of mass velocity are  
\begin{eqnarray} 
v_{cm,max,I} &=& \frac{k}{2} \left(\frac{1}{\gamma_{11}} -\frac{1}{\gamma_2} \right) (h + w) \nonumber \\
v_{cm,min,I} &=& \frac{k}{2} \left(\frac{1}{\gamma_{11}} -\frac{1}{\gamma_2} \right) (h - w).  
\end{eqnarray} 
Similarly, when the dimer is contracting, and $x < l^{\uparrow}$, the maximum and minimum values of the center of mass velocity are given by
\begin{eqnarray} 
v_{cm,max,II} &=& \frac{-k}{2} \left(\frac{1}{\gamma_{11}+\gamma_{12}} -\frac{1}{\gamma_2}\right) (h + w)  \nonumber \\
v_{cm,min,II} &=& \frac{-k}{2} \left(\frac{1}{\gamma_{11} +\gamma_{12}}-\frac{1}{\gamma_2}\right) (h - w).   
\end{eqnarray} 
Finally, the time-averaged-over-one-period $v_{cm}$, or $\bar{v}_{cm}$, is given by 
\begin{equation}
\bar{v}_{cm}=\frac{t_I\bar{v}_{cm,I}+t_{II}\bar{v}_{cm,II}}{t_I+t_{II}},
\end{equation}
where 
\begin{equation}
\bar{v}_{cm,I}=\frac{(\gamma_2-\gamma_{11})}{2t_I(\gamma_{11}+\gamma_2)}(x_I(0)-x_{eq1})(e^{-\frac{k}{\gamma_2}\frac{\gamma_{11}+\gamma_2}{\gamma_{11}}t_I}-1),
\end{equation}
and 
\begin{equation}
\bar{v}_{cm,II}=\frac{(\gamma_2-(\gamma_{11}+\gamma_{12}))}{2t_{II}(\gamma_{11}+\gamma_{12}+\gamma_2)}(x_{II}(0)-(x_{eq1}-x_{eq2}))(e^{-\frac{k}{\gamma_2}\frac{\gamma_{11}+\gamma_{12}+\gamma_2}{\gamma_{11}+\gamma_{12}}t_{II}}-1).
\end{equation}
The time-averaged-over-one-period $v_{cm}$ would presumably be the simplest measurement an experimentalist could perform. So we will study it in detail. The cyclic behavior of $x(t)$ would be more difficult to compare with experiments due to the presence of pseudopods.

Using our parameter estimates from Sec. III, we first present results for $x_{rel}(t)$, $x_{cm}(t)$, and $v_{cm}(t)$. See Figure 4. Apart for the initial cycle, for each subsequent cycle, the time in the extension mode is 5.65 $s$ and the time in the contraction mode is 10.17 $s$. Note that the time scale for the extension mode, which corresponds to the timescale for focal adhesion maturation, is in agreement with the observed timescale of seconds for focal adhesion maturation~\cite{fagrowth}. We find $v_{cm,max,I}=0.088 \,\,\mu m/s$, $v_{cm,min,I}=0.038\,\, \mu m/s$, $v_{cm,max,II}=0.029\,\, \mu m/s$, and $v_{cm,min,II}=0.013\,\, \mu m/s$. The time-averaged-center-of-mass-velocity is $\bar{v}_{cm}=0.033 \,\, \mu m /s$. This value is in reasonable agreement with the order-of-magnitude time-averaged velocity for wild-type HT-1080 fibrosarcoma cells crawling in the ECM~\cite{oscillations}. Of course, we have not yet taken into account the elasticity of the collagen fiber(s) such that we expect our result to be an upper bound on the speed. Interestingly, the maximum instantaneous velocity of the center of mass is the same order as keratocytes crawling on surfaces~\cite{keren}. The time-averaged velocity of the center of mass is about an order of magnitude smaller. So, using physiologically based independent estimates for the parameters involved we obtain reasonable cell speeds for cells traveling in the ECM. 

How does $\bar{v}_{cm}$ vary with the spring parameters, namely, $k$, $h$, and $w$? In Figures 5 and 6, we plot both $\bar{v}_{cm}$ and $x_{cm}(t)$ for several values of these parameters.  As indicated by Eqns (15)-(17), $\bar{v}_{cm}$ increases linearly with the spring constant $k$. On the other hand, increasing the width of the hysteresis loop, $w$, decreases $\bar{v}_{cm}$ since there is a larger strain barrier to overcome to elongate. Once the strain barrier becomes equal to or larger than the added strain energy (due to myosin pulling, for example), i.e. $w>h$, then the active cell can longer move effectively. Moreover, increasing the difference between the two equilibrium spring lengths (increasing $h$), adds more active strain energy into the system with the motors contracting more effectively such that the active dimer can crawl faster until the speed becomes limited by the asymmetry in the friction coefficients. An increase in $h$ can be driven by the addition of myosin (in the contraction mode) or increasing the spring constant associated with the alpha-actinin since the extension mode is driven by releasing strain in the alpha-actinin (as opposed to actin growth).   
\begin{figure}
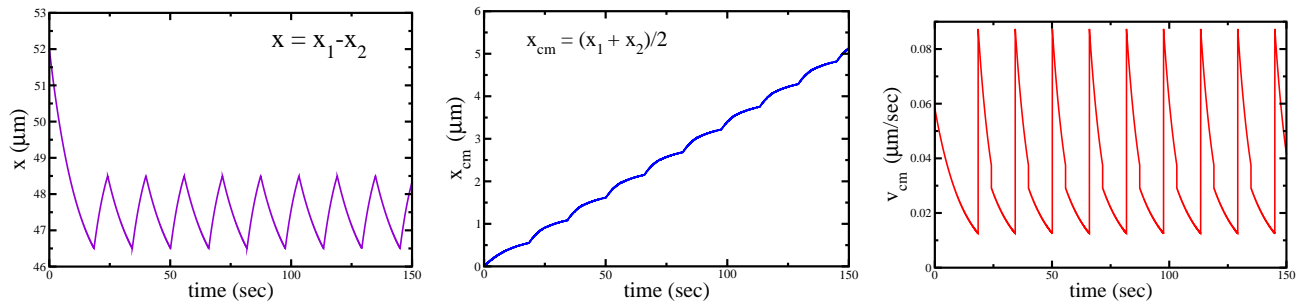

\begin{center}
\includegraphics[width=5.5cm]{xversust.final.eps}\hspace{0.2cm}
\includegraphics[width=5.5cm]{xcmversust.final.eps}\hspace{0.2cm}
\includegraphics[width=5.5cm]{vcmversust.final.eps}

\caption{Left: Plot of cell length $x=x_{1}-x_{2}$ as a function of time for the parameters given in Table I. Center: Plot of position of the center of mass, $x_{cm}$, as a function of time. Right: Plot of velocity of the center of mass, $v_{cm}$, as a function of time. }
\end{center}
\end{figure}

\begin{figure}[h]
\begin{center}
\includegraphics[width=7cm]{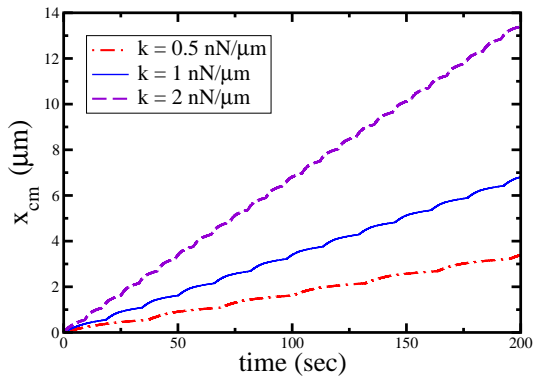}
\caption{Left: Plot of center of mass for cell as a function of time for different spring constants $k$. The parameters are from Table I (unless stated otherwise).}
\end{center}
\end{figure}

\begin{figure}[h]
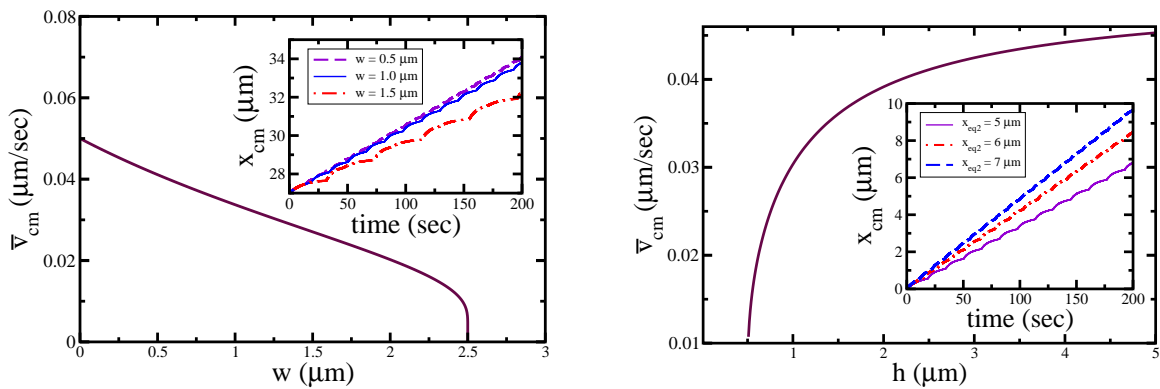

\begin{center}
\includegraphics[width=7.2cm]{width.inset.final.eps}\hspace{1cm}
\includegraphics[width=7cm]{xeq.inset.final.eps}
\caption{Left: Plot of $\bar{v}_{cm}(w)$. Left Inset: Plot of $x_{cm}(t)$ for different widths. Right: Plot of $\bar{v}_{cm}(h)$ for $w=0.5 \,\,\mu m$. Right Inset: Plot of $x_{cm}(t)$ for different heights of hysteresis loop. }
\end{center}
\end{figure}

As stated previously, it is the combination of the nonequilibrium nature of the active spring and the asymmetry of the friction that leads to motion.  We have added this asymmetry explicitly given the molecular understanding of how the integrins behave as catch bonds as focal adhesions mature. In the absence of this asymmetry, i.e. $\gamma_{11}+\gamma_{12}=\gamma_2$ with $\gamma_{12}=0$, then $v_{I,II,cm}=0$.  Moreover, if $\gamma_{12}=0$, then $\bar{v}_{cm}=0$ (even for $\gamma_2\neq\gamma_{11}$) because any new territory gained during the extension mode will be lost during the contraction mode. See Figure 7. Moreover, in breaking the symmetry, we have made a choice as to which direction the active dimer crawls. The cell can change direction when $\gamma_{11}>\gamma_2$ and $\gamma_{12}<0$. Since motion of the center of mass in the extension mode is now to the left, as long as the asymmetry in the friction coefficients in the contraction mode is such that not all new territory gain is lost, then there is net motion to the left. We also observe that as the difference between $\gamma_{11}$ and $\gamma_2$ increases, $\bar{v}_{cm}$ also increases. This increase allows the extension mode of the active dimer to be more efficient at exploring new territory and increases $\bar{v}_{cm}$ (provided $\gamma_{12}\neq 0$ to model the catch bond behavior of the integrin at the front of the cell). See Figure 7.

\begin{figure}[h]
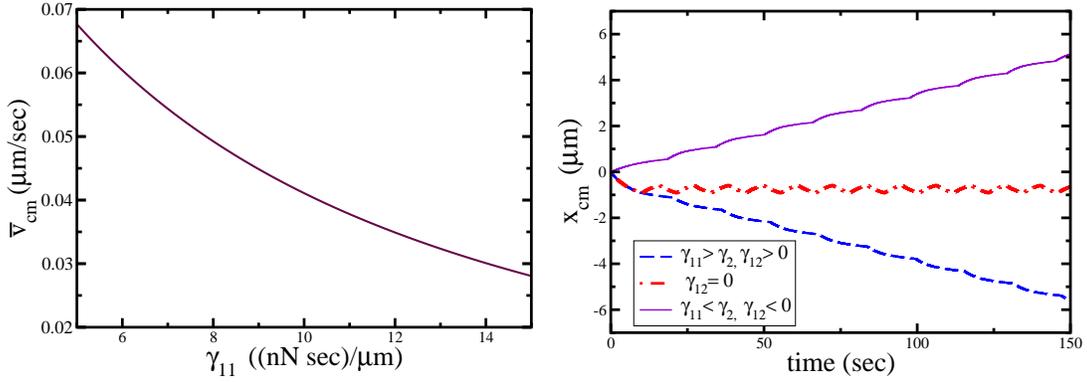

\begin{center}
\includegraphics[width=7cm]{gamma11.final.eps}\hspace{0.2cm}
\includegraphics[width=7cm]{gammas2.final.eps}
\caption{Left: Plot of $x_{cm}(t)$ for different friction coefficients. Right: Plot of $\bar{v}_{cm}(\gamma_{11})$. }
\end{center}
\end{figure}

Now let us investigate the motion of the active dimer when nonequilibrium noise ($A_i>0$) is turned on? Is the motion robust? Why ask this? Well, the cell is very much a dynamic entity. There is mounting evidence that the motion of objects placed in a cell, such as a carbon nanotube, couples to myosin-driven stress fluctuations in the cytoskeleton~\cite{schmidt}. These fluctuations are reminiscient of thermal noise, but with a nonthermal origin. To study the effect of noise on our crawling cell, we simulate the equations of motion using the Euler-Maruyama scheme with $A_i>0$~\cite{euler}. We define $A=A_1=A_2$. 

Given our deterministic active dimer, for small enough values of $A$, the noise can be added perturbatively and should not affect the cyclic behavior of the active dimer. More precisely, we find that for $A < 0.1 \,\, nN^2\,s$, the noise does not affect the motion of the cell with the cyclic behavior between the extension and contraction modes remaining on average (See Figure 8).  However, as $A$ is increased beyond $0.1,\,\,nN^2\,s$, the scallops become washed out on average, though the average speed of the cell remains virtually unchanged.  One can estimate the upper bound of this crossover. When the cell is in the extension mode, for instance, the variance, $\sigma_I(t)$, is given by $\sigma_I(t)=<x_I^2(t)>-<x_I(t)>^2=\frac{A(\gamma_{11}+\gamma_{2})}{k\gamma_2\gamma_{11}}(1-e^{\frac{2k(\gamma_2+\gamma_{11})t}{\gamma_{2}\gamma_11}})$.  When $\sqrt{\sigma_I(t)}$ becomes of order the hysteresis width in the timescale $t_I$ (to use as a first approximation), then the area of the deterministic hysteresis gets washed out on average.  This upper bound corresponds approximately to $A\approx 10\,\,nN^2 s$, which is a bit larger than the observed value. One can improve upon this upper bound by taking into account the directionality of the hysteresis loop and determine the average time scale that the velocity of the relative coordinate goes from positive to negative (a velocity zero-crossing). This is because a velocity zero-crossing can drive the active dimer from one mode to the other.  One can impose a threshold on the noise for this switching to occur.  We leave such modifications for potential future work. What we have learned, however, is that the deterministic model for the model cell is robust to a range of nonequilibrium, or active, noise.  The upper limit of this range maps to an effective diffusion constant of approximately $10^{-3}\frac{\mu m^2}{s}$. 

\begin{figure}[h]
\begin{center}
\includegraphics[width=7cm]{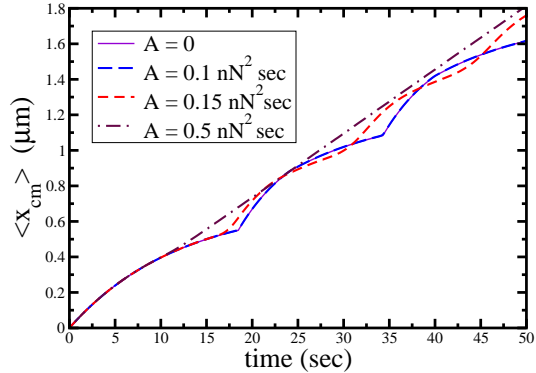}
\caption{Plot of $x_{cm}(t)$ for different values of the noise with $A=A_1=A_2$.}
\end{center}
\end{figure}

Finally, we ask the following question: How does the motion of the active dimer change if the hysteresis loops contain finite slopes?  Then, in going from one mode to the other, the stress fiber would no longer behave as a switch, but the change in equilibrium spring length would depend continuously on the strain. Since the integrins are ultimately coupled to the stress fibers, changes in the friction coefficients would also depend continuously on the strain. Well, as long as curves with finite slope intersect with the $x=x_{eq}$, as is the case with our model, then motion will cease since this is an overdamped system now in equilibrium. See Figure 9. However, the addition of active noise kicks the dimer out of equilibrium and motion resumes. If the active noise is sufficient to change the direction of the strain (extending to compressing, for example), there is a switch from one equilibrium spring constant to the other.  A threshold on this switch will require an active noise strength above this threshold to regain motion. Furthermore, at least for $A_1=A_2$, as the strength of the active noise increases, so does the average velocity of the center-of-mass, or $<v_{cm}>$, though the increasing the active noise strength by an order of magnitude eads to a gain of a few tenths of a percent. In sum, for this finite slope case, active noise is crucial for sustainable net motion.    
\begin{figure}[t]
\begin{center}
\includegraphics[width=7cm]{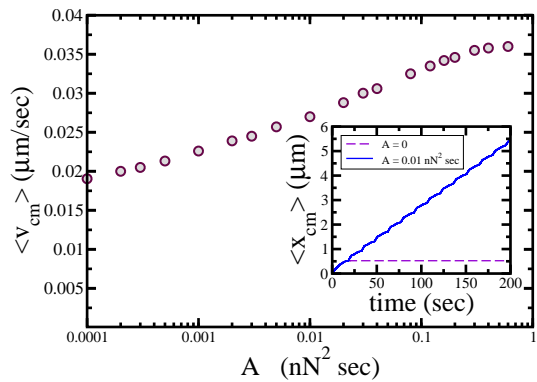}
\caption{Plot of $<v_{cm}>(A)$ for finite slope case with a slope of 5/2. Inset: Plot of $<x_{cm}>(t)$ for $A=0$ and $A=0.01 \,\,nN^2 s$.  }
\end{center}
\end{figure}

\section{Discussion}
We have constructed a minimal model for cell moving on a rigid fiber. The model contains two beads and one spring, the beads representing the front and the back of the cell respectively. Friction coefficients for each bead represent the focal adhesions between the substrate and the cell. We assume the back bead to have constant friction, while the front bead friction changes as nascent focal adhesions become mature focal adhesions to grip the surface via their catch bond behavior. In addition, the single spring connecting the front and the back beads models the basal stress fibers stretching along the cell. The effect of myosin is modeled by a change of the equilibrium spring length. When myosin is actively contracting, the equilibrium spring length is shorter than when myosin is not. We have emphasized that the extension mode is driven by relieving strain in alpha-actinin binding, which could be enhanced due to zyxin binding to alpha-actinin, when the myosin unbind. Both the catch bond behavior and/or dynamics of alpha-actinin may give rise to hysteresis in this active contractility, which we have incorporated into the model. 

We find that the activity of the myosin and the asymmetry in the friction coefficients due to catch bond behavior of the integrins at the front of the cell and slip bond behavior at the back are both needed to obtain directed motion of the crawling cell in an overdamped system in the absence of any noise.   Like Refs.~\cite{recho} and ~\cite{sriram}, our model does not require actin-filament nucleation driven by the branching agent Arp2/3 for cell motility.  This is important for elongated cells crawling along ECM fibers where Arp2/3 plays a role in generating pseudopods to potentially explore new ECM fibers, but does not drive motility~\cite{anjil}. In contrast to Ref.~\cite{recho}, where an advection-diffusion equation for the motor concentration coupled with an active contractile stress drives the motion, our model takes into account the stress fiber structure and the interaction with the subtrate via focal adhesion friction. In constrast to Ref.~\cite{sriram}, our model is deterministic and observes motion in the direction of larger friction (at least for some part of the cycle), which is in keeping with experiments~\cite{dembo}. 

Using independent estimates for the parameters in the model, we find reasonable agreement with observed speeds of elongated cells crawling along ECM fibers~\cite{oscillations}. We also study the average speed as a function of the parameters, which can presumably be qualitatively explored, at least, via knockdowns of the proteins involved or via mutants.  For instance, the larger the difference between the two equilibrium spring lengths, the faster the average cell speed.  A larger difference could be due to more myosin (to enhance the contraction mode), or more alpha-actinin (to enhance the extension mode). Interestingly, increased expression levels of alpha-actinin are found in melanomas and in tumor cell lines with faster migration rates (than the corresponding healthy cells)~\cite{cancer}.  We also find that the net deterministic cell motion is robust to active noise. For the time being, we varied the parameters of the model independently and studied the time-averaged center-of-mass velocity, or speed.  However, varying some of the parameters simultaneously may yield an optimal speed.  

Our model may help understand the finding of oscillations observed in cells that are lacking in the protein zyxin.  More specifically, recent experiments~\cite{oscillations} have found that zyxin-depleted cells migrating in the ECM move persistently along highly linear tracks before reversing their direction. This reversal persists resulting in oscillations. These oscillations have also been observed in cells moving on one-dimensional micro-patterned substrates, but not in two dimensions. Such periodic migration has been shown to result from the coupling between cell shape and actin-polymerization driven polarity in phase-field models of cell migration~\cite{levine}. While protrusive stresses generated by actin filament nucleation via Arp2/3 (and subsequent polymerization) at the leading edge of the cell play a key role in two-dimensional cell migration, it is less dominant in three-dimensional migration. Our model does not require actin filament nucleation and may provide further insight into the underlying mechanism for the above periodic migratory motion in the one and three dimensions. Should zxyin be knocked down, then the switching behavior in our active spring between contraction and extension may become compromised over time (with redundant proteins not as efficient as zxyin) and the cell will eventually not be able to move. Hence, it will fluidize, reorient itself with the help of microtubules, and begin to crawl in another direction to search out new space. In the one-dimensional case, the cell can only reverse its direction to search out ``new'' space. 

One important advantage of our minimal model is that its simplicity easily allows for extension. For instance, we can (1) introduce Arp2/3 generated pseudopods via extra beads and active springs (2) incorporate elasticity into the track, (3) introduce a cell nucleus via extra beads and active springs, and (4) scale up to many cells interacting via cadherins.  As for adding elasticity to the track, the motility of cells migrating in the ECM depends on its microstructure [5-17]. What are, then, the strategies or optimization principles that cells use to migrate in the ECM such that they can harness the elasticity of the ECM fibers to move, while also overcoming the physical barriers to motion imposed by the matrix architecture? We can begin to answer such questions by coupling our model cell to an extensible worm-like polymer and probe the cell's motility. As for introducing a cell nucleus, the discovery of actin stress fibers extending over the nucleus~\cite{actincap} such that as the cell crawls the nucleus is squeezed in the direction transverse to crawling~\cite{Khatau}, begs for study via modeling. We can add these actin cap stress fibers to our basal stress fiber model and address whether their presence helps speed up or slow down a cell crawling along a one-dimensional elastic fiber. And, finally, the extension to interacting active elastic dimers is motivated by recent experiments on a collection of spindle-shaped NIH-3T3 cells at high densities~\cite{silberzan}. Given the geometry of such cells, their mechanism for motion may indeed be similar to one described here. This begs the question, under what conditions does the cell motion {\it not} rely on actin-filament nucleation and polymerization, other than the constrained geometry case of crawling along ECM fibers? Confinement by other cells, potentially of a different type, may indeed be another possibility.

The authors would like to acknowledge helpful discussion with C. Waterman, D. Wirtz, and M. Wu. MD and JMS would also like to acknowledge the hospitality of the Aspen Center of Physics, where part of this work was done.

\end{document}